\documentclass[]{ceurart}

\usepackage{listings}

\begin{document}

\copyrightyear{2022}
\copyrightclause{Copyright for this paper by its authors.
  Use permitted under Creative Commons License Attribution 4.0
  International (CC BY 4.0).}

\conference{IAIL 2026: Imagining the AI Landscape after the AI Act, Brussels, Belgium, July 7, 2026}

\title{Lost in Vagueness: Towards Context-Sensitive Standards for Robustness Assessment under the EU AI Act}

\author[1]{Roberta Tamponi}[email=r.tamponi@liacs.leidenuniv.nl]
\author[2,3]{Carina Prunkl}
\author[1]{Thomas B\"ack}
\author[1]{Anna V.~Kononova}
\address[1]{Leiden Institute of Advanced Computer Science (LIACS), Leiden University}
\address[2]{University of Oxford}
\address[3]{Inria}

\begin{abstract}
Robustness is a key requirement for high-risk AI systems under the EU Artificial Intelligence Act (AI Act). However, both its definition and the methodologies for assessing it remain underspecified, leaving providers with little concrete direction on how to demonstrate compliance. This stems from the Act’s horizontal approach, which establishes general obligations applicable across all AI systems, but leaves the task of providing technical guidance to harmonised standards. This paper investigates what it means for AI systems to be \textit{robust} and illustrates the need for context-sensitive standardisation. We argue that robustness is not a fixed property of a system, but depends on which aspects of performance are expected to remain stable (“robustness of what”), the perturbations the system must withstand (“robustness to what”) and the operational environment. We identify three contextual drivers—use case, data and model—that shape the relevant perturbations and influence the choice of tests, metrics and benchmarks used to evaluate robustness. The need to provide at least a range of technical options that providers can assess and implement in light of the system’s purpose is explicitly recognised by the standardisation request for the AI Act, but planned standards, still focused on horizontal coverage, do not yet offer this level of detail. Building on this, we propose a multi-layered standardisation framework that embeds context sensitivity in the ongoing standardisation process. Horizontal standards define shared principles and terminology, while domain-specific ones identify relevant risks and map them across the AI lifecycle, guiding the selection of best practices for specific applications. These, together with benchmarks and sandboxes, should be organised in a dynamic repository where providers can propose new informative methods and share lessons learned. Such a system reduces the interpretative burden, mitigates arbitrariness and addresses the obsolescence of static standards, ensuring that robustness assessment is both adaptable and operationally meaningful.
\end{abstract}


\maketitle

\section{Introduction}
In 2024, the European Union (EU) adopted the Artificial Intelligence Act (AI Act), the world's first comprehensive legal framework for AI technologies~\cite{act2024artificial}. 
A central challenge is that, despite outlining procedural obligations, the AI Act offers little detail on how these should be put into practice~\cite{floridi2022capai}. Criticisms were also raised for the imprecision of some of its key terminology, which further complicates its practical implementation~\cite{finck2026eu, nolte2025robustness, deck2024implications, laux2024trustworthy}. This lack of clarity reflects the Act’s horizontal approach, which sets general requirements so that they can apply across diverse AI systems, making some degree of abstraction inevitable. Aware of this, the AI Act’s implementation is meant to be supported by harmonised standards which are intended to provide the technical specifications needed for the design, development and conformity assessment of high-risk AI systems~\cite{EU2023StandardisationAI}.

To examine how the AI Act’s requirements can be operationalised in practice, this paper focuses on the \textit{robustness} requirement, given the limited attention it has received in the compliance literature~\cite{nolte2025robustness} and its strong dependence on context~\cite{freiesleben2023beyond}. 
Art. 15 requires high-risk AI systems to achieve an appropriate level of accuracy, robustness and cybersecurity throughout their lifecycle and to be resilient to errors, faults, inconsistencies and unexpected situations. While these provisions set out useful normative requirements, they do not provide a formal definition of robustness and remain too open-textured to be operationalised without additional specifications. A similar pattern appears in the literature, where researchers either focus on narrow technical definitions or leave the concept too abstract, aiming for generality. This ambiguity reflects the concept's \textit{context-sensitive nature}: the robustness of a system depends on its intended purpose, design and deployment environment, which shape the perturbations and potential failures most relevant for evaluation. 
Hence, practical assessment is context-sensitive, since even minor changes in operating conditions may require different tests, metrics, thresholds and mitigation strategies. 
Although the need to account for context is recognised both by the AI Act and by the standardisation request~\cite{EU2023StandardisationAI}, current standardisation efforts remain largely oriented towards horizontal coverage and offer limited support for operationalising this requirement across different contexts. The only international standard on robustness adopted so far by the EU, the \citet{iso2021artificial}~\footnote{
“CEN/CLC ISO/IEC” indicates that the technical report (TR) by ISO (International Organisation for Standardisation) and IEC (International Electrotechnical Commission) has been adopted as a European standard.}, provides only an overview of the assessment methods for neural networks. 
Other standards on robustness are under development, but their specificity remains uncertain. 
The risk of overly generic guidance, therefore, remains, leaving providers with excessive interpretative burden and the risk of arbitrary implementation.

In this interdisciplinary paper, we aim to bridge regulatory interpretation and practical evaluation by examining what it would take to operationalise the robustness requirement under the AI Act. We first address the term's ambiguity by proposing a clearer conceptualisation of robustness that distinguishes the \textit{robustness of what}, that is, the aspects of system performance expected to remain stable, from \textit{the robustness to what}, which refers to the perturbations the system must withstand. 
We identify \textit{use case}, \textit{data} and \textit{model} as the contextual drivers that determine the relevant perturbations and guide the choice of the most suitable methods for assessing robustness. 
This analysis reveals a second gap: although robustness assessment is inherently context-sensitive, current standardisation efforts do not yet provide guidance on how to make context-dependent choices.
To overcome this limitation, we propose a multi-layered framework to embed context sensitivity into the ongoing standardisation process. The framework combines horizontal and domain-specific standards. For each domain, it includes a context-sensitive perturbation taxonomy that maps the main risks to robustness arising from use case, data and model. To address the limited adaptability of standards to technological change, the framework includes a dynamic repository of best practices to test, measure and mitigate perturbations, as well as benchmarks and sandboxes. Through this repository, providers can also propose new informative methodologies and share their experiences. 

The paper proceeds as follows. Section~\ref{sec2} outlines the AI Act and its treatment of robustness.
Section~\ref{sec3} highlights the limits of overly abstract definitions of the term. Section~\ref{sec4} distinguishes the two dimensions of robustness, namely the robustness of what and the robustness to what. 
Section~\ref{sec5} defines the contextual drivers, examines how they shape relevant perturbations, analyses adversarial and distribution shift perturbations, discusses the context sensitivity of the assessment methods and concludes with a use case comparison. Section~\ref{sec6} discusses current standardisation challenges and Section~\ref{sec7} proposes the multi-layered framework.

\section{Robustness in the AI Act} \label{sec2}

Adopted in 2024, the AI Act regulates the development, the placing on the market and the use of AI systems in the EU, while promoting the uptake of human-centric and trustworthy AI. It establishes a product-safety-based regulatory framework that applies different obligations depending on the risk level of AI systems. \textit{Risk} is defined as the combination of the probability and severity of harm (Art. 3) and systems are classified into four categories: \textit{unacceptable, high, limited,} and \textit{minimal} risk. The AI Act primarily targets high-risk systems, those that can significantly affect safety, fundamental rights, or critical domains.  
These must obtain the "CE" marking (Art. 48) through a conformity assessment (Art. 43, Annex VI–VII), demonstrating compliance with Chapter III, Section 2, which sets requirements on risk management (Art. 9), data governance and quality (Art. 10), record keeping (Art. 12), transparency (Art. 13), human oversight (Art. 14), accuracy, robustness and cybersecurity (Art. 15). Providers must maintain technical documentation (Art. 11), implement a quality management system (Art. 17) and, after deployment, monitor performance using tools defined during the development. Some provisions have taken effect since 2025, while others will become applicable gradually up to August 2027, allowing time for the development of harmonised standards and the establishment of supervisory authorities. 

Robustness is a core requirement. Article 15 states that "high-risk AI systems shall be designed and developed in such a way that they achieve an appropriate level of accuracy, robustness, and cybersecurity, and that they perform consistently in those respects throughout their lifecycle" (Art. 15.1). Article 15.4 specifies that these systems "shall be as resilient as possible regarding errors, faults or inconsistencies that may occur within the system or the environment in which the system operates", it also states that robustness may be achieved through technical redundancy solutions, including backup or fail-safe plans and requires that systems that continue to learn after placement on the market include mitigation measures to minimise the risk of biased outputs arising from feedback loops. Recital 75 clarifies that robustness is intended to prevent harmful or undesirable behaviour resulting from system limitations or environmental conditions and emphasises that such failures may have safety implications or affect fundamental rights.

\citet{finck2026eu} argues that Article 15 is filled with ambiguous legal concepts and vague obligations, making its interpretation and implementation challenging. Formulations such as "appropriate level" and "as resilient as possible" leave room for differing interpretations of what constitutes sufficient protection and, despite their technical appearance, these requirements involve normative judgements yet provide limited guidance on how to demonstrate compliance in practice, underlining the future importance of harmonised standards. 
This challenge is intensified by the probabilistic nature of many AI systems, whose outputs are shaped by statistical uncertainty, while compliance assessment usually relies on fixed, measurable criteria. 
As she observes, "without clear predefined benchmarks, providers must interpret and implement measures based on their own assessment, potentially leading to inconsistency in compliance approaches"~\cite{finck2026eu}. 
This interpretative uncertainty is compounded by the heterogeneity of AI systems. As noted by~\citet{malgieri2026eu}, AI applications differ in purpose, design and operational context, requiring case-by-case measures, thereby placing a considerable interpretative burden on providers. While this flexibility is consistent with the risk-based approach of the AI Act, it also complicates the translation of general requirements into concrete and verifiable compliance practices. Other concerns have been raised by~\citet{nolte2025robustness} who notice that Article 15 also leaves unclear how robustness should be measured in relation to accuracy, what it means for performance to remain “consistent” throughout the system’s “lifecycle” and even what level of robustness is required, given the tension between the “appropriate level” (Art. 15.1) and the stronger “as resilient as possible” formulation (Art. 15.4). They further observe that the Act uses the terminology of “robustness” and “technical robustness” inconsistently and does not fully reflect the taxonomy of the machine learning literature, as adversarial robustness appears to be addressed primarily under Article 15.5, which concerns cybersecurity.

Notably, the AI Act does not offer a formal definition of robustness. 
Under the EU’s drafting guidance~\cite{european2003joint}, where a term is "not unambiguous", it should be defined in a single definitions article at the beginning of the act and definitions should not contain autonomous normative provisions. Yet robustness does not appear among the definitions in Article 3 and Article 15 only sets out general obligations. Unambiguous terms would not need to be defined in the regulation, but as prior literature proves, this is not the case of robustness~\cite{finck2026eu, nolte2025robustness, tocchetti2025ai, freiesleben2023beyond}. The regulation itself acknowledges that more specifications are needed. Article 15.2 recognises that benchmarks and measurement methodologies are still lacking and indicates that the Commission is working with stakeholders to address the technical aspects of how to measure the appropriate levels of robustness.
This is coherent with the aim to create a horizontal regulation that applies to all AI systems, but highlights the need for complementary detailed standards that include technical specifications. 
Yet, despite the ongoing standardisation process, major uncertainties remain about the level of detail these standards will provide and, in their current form, both the definition of robustness and the related provisions remain insufficiently specified.

\section{Limitations of an abstract definition of robustness} \label{sec3}

The notion of robustness in AI remains conceptually ambiguous~\cite{tocchetti2025ai, freiesleben2023beyond}. To unpack what it means for a system to be robust, it is useful to begin with a definition of the concept. Since the AI Act does not provide one, a natural starting point is the existing harmonised standard on the robustness assessment of neural networks \citep{iso2021artificial}, which defines robustness as the ability of an AI system to \textit{maintain its level of performance under any circumstances}. While this definition may be adopted for regulatory purposes, such as the AI Act, access to standards is restricted by copyright and they are not publicly available. As a result, researchers often rely on alternative definitions. For example,~\citet{nobandegani2019robustness} define it as the insensitivity of a model’s performance to miscalculations of its parameters. \citet{freiesleben2023beyond} describe it as a model's capacity to sustain stable predictive performance in the face of variations in input data. Across subfields, the term has taken on multiple meanings. In computer vision, the term has been used to denote raw performance on held-out test sets, generalisation within and across domains, maintaining performance on manipulated inputs and naturally-induced image corruptions and resistance to adversarial attacks~\cite{drenkow2021systematic}. In natural language processing (NLP), \citet{la2020assessing} frame robustness of text classification in terms of the capacity to maintain predictions under word substitutions, formalised as a maximal safe radius for a given input text.

Despite the variety of formulations, the underlying principle to which they all refer is \textit{the stability of performance}. This broad principle can be captured in the high-level definition: \textit{a model is robust when small changes in input cause small changes in output}. While this is an intuitive starting point, it remains too abstract to be operationally useful. The definition does not specify both the nature of the changes in input and to which output it refers. A more comprehensive definition might be: \textit{a system is considered robust to some kind of small perturbations in the input data when the system's performance under the perturbed state remains similar to that of the unperturbed state}. 
However, the relevant perturbations, as well as what types of performance remain underspecified. The necessary clarifications depend on the context in which the system is developed and deployed, which is why subfield case-specific definitions of robustness coexist in the literature. \textit{Robustness is not a fixed notion of a system, but depends on the system's different assumptions and goals}. As a result, a system considered robust in one context may not be robust in another. 

\section{Robustness dimensions} \label{sec4}

Building on this context-sensitive understanding, this section proposes a structured way of interpreting robustness that supports practical evaluation. In line with~\citet{freiesleben2023beyond}, who frame robustness as a relation between targets, modifiers, domains and tolerance thresholds, we share the view that robustness must be context-specific. Our focus, however, is on linking abstract definitions to practical evaluation. 
Starting from the principle of the stability of performance, we distinguish between two core aspects: the system's output that is required to remain stable (\textit{robustness of what}) and the types of changes in input it is required to withstand (\textit{robustness to what}). 

\subsection{Robustness of what?}

To assess robustness in practice, what is expected to remain stable must first be specified. In machine learning, it is the system’s \textit{performance}, as robustness concerns the system’s ability to maintain adequate predictive capabilities under non-ideal or varying conditions. However, performance itself is neither fixed nor universal: what counts as adequate performance is subject to interpretation and depends on the system’s purpose and operational context. It may refer to accuracy~\cite{dong2020benchmarking}, runtime~\cite{singh2019boosting}, precision~\cite{yang2021enhancing}, etc. 
In line with the terminology used in current international standardisation efforts, \citet{2024BEI2} reports that \textit{performance} in the AI field refers to \textit{how well a certain AI system performs the intended tasks}. Performance can be measured with relevant metrics that depend on the system's purpose. \citet{iso2021artificial} proposes an exhaustive list which includes accuracy, precision, recall, specificity, $F_1 $ score, ROC and AUC. These can be used independently or in combination, but they are not interchangeable. Notably, there are also numerous task-specific metrics, such as BLEU, TER and METEOR for machine translation, or mean average precision for ranked retrieval. 
To be meaningful indicators of performance, metrics need to be chosen based on the system's purpose and the impact of the different types of errors. 
For example, in a child-safety video filter, where the retrieved set is the non-harmful content, high precision is preferred:  excluding some acceptable videos (low recall) is better than showing harmful content. In shoplifter detection, on the other hand and depending on operational preferences, false alerts may be tolerated if this ensures that most shoplifters are detected, so recall may be prioritised over precision~\cite{geron2022hands}.  
Since such measures often trade off, the system's function and associated risks must be understood before identifying which aspects of performance are most crucial to maintain under perturbations.
Robustness evaluation assumes a fully trained and tested system with established performance under standard conditions.
This baseline is then used to assess how perturbations affect performance and to define tolerable deviation thresholds. These thresholds depend on deployment context, with a lower tolerance level due to higher associated risks  (e.g., clinical decision-making).

\subsection{Robustness to what?}

After clarifying the role of performance in robustness evaluation, the next step is to identify against what the model must be robust. Robustness assessment concerns how input perturbations affect the system's predictive capabilities and whether the model can maintain its intended behaviour. 
There are several possible threats to an AI system. As factors that challenge robustness, the~\citet{2024BEI2} identifies the presence of unseen, biased, adversarial or invalid data inputs, external interference and environmental conditions. Similarly, the AI Act identifies errors, faults, inconsistencies and unexpected situations (Recital 75, Art 15.4). 
While these risk factors may appear relatively specific, they remain difficult to operationalise, as they group together different phenomena and do not indicate how their impact should be assessed or prioritised in context. They can therefore be understood as high-level sources of risk, which require further specification to be translated into testable perturbations.
Some minor perturbations in input are unavoidable, while others might lead to more severe problems. A report of the latter must be documented under the risk management obligations of Art. 9, which requires the identification, estimation and evaluation of foreseeable risks. 
Identifying which input changes are critical depends on the system’s intended use; therefore, the robustness goal must be clearly specified. Generic instructions, such as “robust to dissimilar inputs”, are insufficiently precise and depending on the type of variation, a system can comply or behave inadequately. 
A system may maintain stable performance on data that presents a gradual shift in the distribution, but can fail if the inputs belong to edge-case scenarios that were insufficiently represented during training.
Assessing robustness requires a precise goal, defined by narrowing down the relevant perturbations that the system can face. The following section introduces the contextual drivers and discusses how they shape both the identification of possible perturbations and the choice of evaluation methods.

\section{Context sensitivity of robustness assessment} \label{sec5}

\begin{figure*}[t]
    \centering
    \includegraphics[width=1\textwidth]{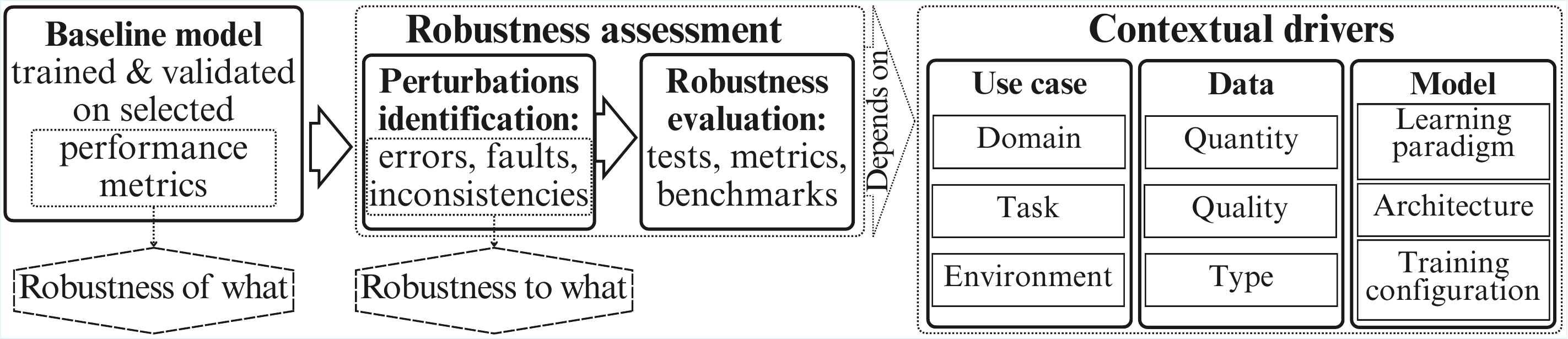} 
    \caption{Conceptual pipeline for context-sensitive robustness assessment.}
    \label{fig:framework}
\end{figure*}

Not all perturbations are equally relevant for every system, the most critical ones must be identified in light of the intended application, system design and deployment environment. For self-driving cars, protecting against adversarial attacks is crucial, but it is not necessarily relevant for an epidemiologist using AI models to forecast virus spread~\cite{freiesleben2023beyond}. 
We identify three contextual drivers—use case, data and model—that guide the selection of relevant perturbations and, in turn, inform the tests, metrics and benchmarks through which robustness is evaluated in a context-sensitive way. Figure ~\ref{fig:framework} provides a schematic overview of context-sensitive robustness assessment.

\subsection{Contextual drivers of robustness} \label{sec5.1}

\paragraph{Use case. } 

When developing an AI system, its intended use case needs to be specified. This includes its application domain, the task and the deployment environment in which it is expected to operate. 
The AI Act (Annexes I and III) gives several examples of domains where AI use can be considered high-risk, such as healthcare, education, employment, access to services, law enforcement, migration and administration of justice. 
Within each domain, AI may serve different tasks, for example, real-time health monitoring, school grading, needs-based resource allocation, etc. The deployment environment refers to the real-world conditions in which the AI system is expected to operate, which is typically dynamic, partially observable and/or subject to noise or user interaction. Addressing these different aspects means delineating the problem that the AI system is meant to solve and how, by clarifying the goals and the requirements that need to be met~\cite{floridi2022capai, chollet2021deep}. 
In healthcare, for example, systems may require robustness to co-occurring conditions such as gastrointestinal or neurological disorders, or robustness to variations in computed tomography protocols, including scanner model, patient size, or radiation dose~\cite{balendran2025scoping}. 
Understanding the use case helps identify the system's potential failures and limitations, guiding both the perturbations to be addressed and the criteria for assessing them.

\paragraph{Data. } 

Data quantity, quality and type determine applicable models, relevant perturbations and suitable evaluation metrics. A main challenge while building an AI system is the lack of training data: even basic problems need thousands of examples, while complex ones may require millions ~\cite{geron2022hands}.  Shallow learning models like decision trees or support vector machines often work well with limited data, while neural networks require more data. Another challenge is the presence of non-representative and poor-quality data. A system can learn and detect underlying patterns only if the training data include enough relevant features that are not full of errors, outliers and noise~\cite{geron2022hands}. 
Robustness to data drift, for example, depends on how performance degrades due to shifts in the patterns or in the environment~\cite{quinonero2022dataset}. Low-quality data are detrimental to adversarial training, as they cause overfitting and overestimation of robustness against weaker adversaries such as projected gradient descent (PGD)~\cite{dong2021data}. The use of large and diverse datasets increases generalisability, making a system more robust to variations and noise sources in data~\cite{javed2025robustness}. Finally, the data type: even when robustness is framed in terms of distribution shift, its manifestation is data-specific. In images, perturbations may appear as noise, blur or weather effects, each testable through specific techniques. For text they can be probed at character, word, or sentence level, reflecting typos, or stylistic changes, each requiring tailored evaluation methods \citep{qiu2022benchmarking}.

\paragraph{Model. }

Different models have different sensitivities to data variations and robustness depends on learning paradigm, architecture and training configuration choices. 
The learning paradigm is the way in which a model learns from data. 
For example, in quality assurance, image anomaly detection can be addressed with models based on different learning paradigms, including supervised and unsupervised deep learning, with the latter being potentially more effective when only a small amount of anomalous data is available, or when the data are unlabeled~\cite{wilmet2021comparison}.
The architecture refers to the different model families. Linear classifiers like Naive Bayes may be preferred for their efficiency and interpretability, but where systems must handle non-linear perturbations, models such as k-Nearest Neighbours might be more appropriate~\cite{christopher2008introduction}.
High-risk applications increasingly rely on foundation models, including large language models (LLMs). LLMs' robustness must be evaluated with respect to behavioural consistency under prompt perturbations, long-context noise, ability to detect and handle out-of-distribution inputs and hallucinations ~\cite{zhang2025evaluating}. 
Moreover, some models' architectures can better mitigate the accuracy/robustness trade-off than others, balancing the performance on clean data with the architecture’s ability to contrast adversarial perturbations~\cite{deng2019architecture, tsipras2018robustness}. Training configuration also has to balance this trade-off. For example, the inclusion of batch normalisation has been shown to improve model accuracy, but in some models, it increases the vulnerability to adversarial attacks~\cite{juraev2024impact, ghaffari2022adversarial}. Hyperparameters such as learning rate, batch size and regularisation coefficients are likewise critical for robustness and optimisation strategies like grid or random search can substantially enhance it by fine-tuning these settings  \citep{javed2025robustness, roy2023hyperparameter, arnold2024role}. Training configuration choices of LLMs include the use of vast and heterogeneous pre-training corpora, prompt-based task unification, parameter-efficient fine-tuning strategies and alignment procedures based on human feedback, which differ from classical full-parameter tuning and directly affect model stability~\cite{zhang2025evaluating}.

\subsection{Robustness evaluation} \label{sec5.2}

Use case, data and model can guide the selection of relevant perturbations, which in turn define the tests, metrics and benchmarks used to assess robustness. 
Relevant perturbations cannot be defined unambiguously, as the deployment environment can introduce unpredictable forms of uncertainty that are difficult to anticipate or quantify. The literature commonly organises robustness into two macro-categories: adversarial robustness and robustness to distribution shift. 

\textit{Adversarial robustness} is the ability of a model to maintain stable performance under intentional manipulations of the input data, designed to mislead it~\cite{zheng2021graph}. These perturbations are non-random and precisely constructed to subtly change the input and maximise the probability that the model will produce an incorrect prediction~\cite{szegedy2013intriguing}. Adversarial perturbations are commonly categorised by the attacker’s knowledge of the target model, typically distinguishing between white-box and black-box settings. In white-box scenarios, the attacker has direct access to the model architecture, parameters, or training data, while in black-box settings, the attacker can only observe the model's responses to certain inputs. 
 Another categorisation, useful for assessing both the feasibility and the potential impact of an attack,  is drawn between targeted and untargeted attacks. The first aims to induce a specific incorrect prediction chosen by the attacker, while the latter is indifferent to the misclassification produced as long as the prediction is incorrect~\cite{braiek2025machine}.  In the healthcare domain, \citet{balendran2025scoping} document that adversarial attacks have been studied in relation to the manipulation of clinical text, the alteration of diagnostic imaging and the perturbation of physiological signals.

\textit{Robustness to distribution shift}, also discussed in the literature as \textit{out-of-distribution (OOD) generalisation}~\cite{hendrycks2021many}, concerns the robustness to perturbations that occur spontaneously in the deployment environment, which introduce a deviation between the distribution of the test data and the one on which the model was trained~\cite{koh2021wilds, drenkow2021systematic}. However, this standard definition has been criticised as overly general and uninformative. \citet{farquhar2022out} show that OOD refers to four qualitatively different scenarios that include transformed distributions, which refer to changes from the original distribution by some set of transformations, related distributions, where distributions are similar to the original in a way determined by the application context, complement distributions, which involve inputs that lie outside the support of the original data distribution and synthetic distributions that are easy to sample but unrelated to the original. Beyond differences in kind, \citet{li2023robustness} argue that robustness to distribution shifts can vary substantially depending on the degree of the shift and demonstrate that model behaviour under mild shifts is often a poor predictor of its behaviour under more severe shifts. \citet{braiek2025machine} emphasise that distribution shift perturbations are rarely characterised and identifies the mechanisms through which they arise. These mechanisms include environmental conditions, such as changes in illumination or weather, and gradual data drift driven by external dynamics like evolving social behaviours or economic trends. Performance may also deteriorate if models rely on spurious correlations or irrelevant patterns present in training but absent at deployment, or encounter rare or edge-case scenarios that were insufficiently represented during training.  
In healthcare, robustness to distribution shift can be challenged by shifts across care settings, demographic subgroups and technical variations in imaging protocols, such as differences in scanner models, acquisition parameters, or imaging views~\cite{balendran2025scoping}. This example highlights how these perturbations are deeply context-sensitive.

Once the relevant perturbations that may pose a risk to the system have been identified, they have to be tested to determine whether the system can remain functional.
To test robustness means submitting the model to various types of perturbations to identify vulnerabilities and weaknesses~\cite{lassance2019structural}.  Various testing techniques have been investigated. For adversarial attacks, for example, generative adversarial networks (GANs)~\cite{goodfellow2014generative}  can be used to synthesise challenging test cases. In the metamorphic testing~\cite{tian2018deeptest,zhou2019metamorphic}, the model's behaviour is evaluated under meaningful input transformations. Mutation testing~\cite{shen2018munn, ma2018deepmutation} introduces small faults into the test data to evaluate whether the model can detect them.
To evaluate the test, metrics are needed to quantify how much a model's performance degrades when exposed to perturbations. \citet{iso2021artificial} overview on the robustness of neural networks assessment distinguishes \textit{statistical methods}, where robustness is defined as a limited performance drop relative to an unperturbed reference set, \textit{formal methods}, which provide mathematical proofs but rely on restrictive assumptions, and \textit{empirical methods}, which rely on experimentation, observation and expert judgement.  Benchmarking an AI system can also help assess some degree of its robustness, as it provides shared and standardised conditions to compare models under controlled perturbations in a reproducible way.  
The AI Act encourages the development of benchmarks to support the evaluation of high-risk AI systems, but leaves open what counts as an appropriate benchmark and who should be responsible for their creation or maintenance. 
Some benchmarks focus on specific types of robustness: RobustBench standardises adversarial robustness evaluation across models~\cite{croce2020robustbench}, while WILDS targets robustness to distribution shifts, including domain generalisation and subpopulation shift~\cite{koh2021wilds}. Some are tailored to a learning paradigm: OpenAI Gym~\cite{brockman2016openai} supports robustness testing in reinforcement learning. While others focus on optimisation methods, for example, in heuristic optimisation, IOHprofiler evaluates algorithm selection and configuration under perturbed scenarios~\cite{doerr2018iohprofiler, wang2022iohanalyzer}.
Benchmarks can thus help establish initial trust in an AI solution even before deployment, but their relevance depends on the type of robustness being tested and the task at hand. 

\subsection{Comparing use cases } \label{sec5.3}

\begin{table}[t]
\centering
\small
\renewcommand{\arraystretch}{1.3}
\caption{Comparison of two use cases where changes in the design choices lead to different perturbations and therefore different methods for robustness evaluation.}

\begin{tabular}{p{1.5cm} p{7.6cm} p{5.0cm}}

\toprule
    \textbf{Details} & \textbf{Use case 1} & \textbf{Use case 2} \\

\midrule
    Use case & Detection of renal cell carcinoma (RCC) & Detection of tuberculosis-related features   \\ \hline
    Data & Whole-Slide Images (WSIs) of Hematoxylin and Eosin (H\&E) stained tissue, different datasets for training and testing & Two datasets of CXR: (1) confirmed tuberculosis, (2) tuberculosis-associated features  \\ \hline
    Model & CNN (ResNet) & CNN (Inception V3) \\ \hline
    Perturbation& Adversarial attack & Distribution shift   \\ \hline
    Robustness test & 1) Multiple adversarial attacks (FGSM, PGD, FAB, Square Attacks, AutoAttack, AdvDrop), 3 different attack strengths,
2) Adversarially robust training, 
3) Dual Batch Normalisation training,
4) Selection of 450 tiles from the RCC set to test the degree of misclassification &  Train the model with one dataset and test it on one with a different distribution\\ \hline
    Metric & 1) AUROC before-after perturbations 2) ASR & AUROC before and after distribution shift  \\

\bottomrule
\end{tabular}

\label{table2}
\end{table}

Even within similar contexts, what makes a system robust may change drastically and, with it, its assessment. Table~\ref{table2} compares two studies on AI-based medical image classification for disease detection. The first investigates convolutional neural networks (CNN) classifiers for renal cell carcinoma (RCC)~\cite{ghaffari2022adversarial}. The second examines CNN for tuberculosis-related chest radiographs (CXR) classification~\cite{sathitratanacheewin2020deep}. Use case 1 trains on data from the Cancer Genome Atlas and tests on data from the University of Aachen, while use case 2 trains on CXR from Shenzhen Hospital in China and tests on both a held-out Shenzhen set and the NIH ChestX-ray8 dataset from the USA, which includes tuberculosis-related features but also other lung abnormalities. Robustness in case 1 is tested against six adversarial attacks that differ in nature and strength (white-box vs. black-box, ensemble and transformation-based attacks),  complemented by adversarial training, dual batch normalisation and evaluation of attack success rates (ASR) on selected image subsets, while, in case 2, it is assessed through domain shift, testing cross-population generalisability from Chinese to USA data. In both cases, robustness is measured by comparing AUROC (ability to discriminate between true and false positives) before and after perturbations, with use case 1 also reporting ASR to capture the degree of misclassification under adversarial inputs.

This comparison shows that even within similar contexts, disease detection through CNN on image data, what defines the system as robust can change significantly. Small shifts in the assumption, in this case regarding the deployment environment, lead to very different evaluations. In one case, the deployment environment involves controlled digital pathology workflows where a plausible risk is the presence of adversarial manipulations. Therefore, robustness must be tested against gradient-based perturbations, minimal decision-boundary changes, random or combined attacks and high-frequency manipulations. In the other case, the deployment environment is expected to include data from different populations; here, the perturbation is distribution drift and the model has to be robust to changes in the data distribution between training and deployment. Similar variation can also occur within the same class of perturbations as a result of changes of different contextual drivers: for instance, in large language models for medical question-answering, adversarial robustness can be tested through MedFuzz attacks that introduce misleading patient characteristics~\cite{ness2024medfuzz}.

\section{Current standardisation directions} \label{sec6}

Harmonised standards may be used to demonstrate compliance with the AI Act and, while not mandatory, they grant the presumption of conformity (Art. 40) and are fundamental for its practical implementation. 
The European Commission is working with the European Standardisation Organisations (ESOs), which include the European Committee for Standardisation (CEN), the European Committee for Electrotechnical Standardisation (CENELEC) and the European Telecommunications Standards Institute (ETSI), to develop technical specifications for the AI Act. In May 2023, CEN and CENELEC accepted the standardisation request M/593. The JTC 21 (Joint Technical Committee), established by CEN and CENELEC, is the dedicated body developing standards in support of the Act, complementing and adapting existing international standards to the Act's requirements. At the time of writing, its work programme includes 34 ongoing activities and 17 published standards~\footnote{See the CEN/CLC/JTC 21 \href{https://standards.cencenelec.eu/ords/f?p=205:22:::::FSP_ORG_ID,FSP_LANG_ID:2916257,25&cs=114251C6C0B684FBBC069923513BF6348}{\textbf{Work Programme}} and \href{https://standards.cencenelec.eu/ords/f?p=205:32:::::FSP_ORG_ID,FSP_LANG_ID:2916257,25&cs=114251C6C0B684FBBC069923513BF6348}{\textbf{Published Standards}}.}, including both CEN/CENELEC developments and ISO/IEC standards adopted as European ones. The deliverables were expected in 2025, but completion is now expected in 2026~\cite{EUCommission2026UnderstandingStandardisationAIAct}. 

The Joint Research Centre  (JRC) report \textit{Harmonised Standards for the European AI Act} emphasises that standards should consist of a small number of horizontal requirements, applicable across different AI systems and sectors, complemented by sector-specific provisions only when strictly necessary~\cite{JRC139430}. On the other hand, the report expects standards to be “sufficiently prescriptive and clear” to support practical application for specific use and the identified risks, guiding providers in selecting appropriate tests and measures. The report thus expects standards to be both broadly applicable and context-sensitive, a hardly feasible ambition. 
The AI Act itself recognises the need for context sensitivity. Article 15 refers to an “appropriate” level of robustness and more generally ties obligations to the system’s intended purpose, which suggests that robustness cannot be treated as a uniform requirement across all AI systems and raises the question of whether horizontal standards can concretely capture context sensitivity. 
The \citet{EU2023StandardisationAI} is more straightforward about what European standards should deliver. The request calls for European standards to take into account the risks common (horizontal) to AI systems in general. Yet, notwithstanding their horizontal nature, it also makes clear that standards may and, where relevant, should provide specifications for particular domains and use cases, taking into account the intended purpose and context of use of those systems. Standards must also reflect the acknowledged state of the art and provide concrete, verifiable technical specifications, including design and development requirements, as well as verification, validation and testing procedures. Even if not every specific intended purpose can be covered, standards are expected to set out at least a range of technical options that providers can assess and implement in light of their system’s purpose, together with guidance on how such options should be applied. The request thus explicitly acknowledges the need for a context-sensitive approach, making it difficult to argue that horizontal standards alone could fulfil the mandate. Horizontal standardisation may support efficiency and maintainability, but when the requirements vary significantly with context, as in the case of robustness, it risks becoming overly generic and detached from practice. 

The current work on robustness standards focuses on horizontal coverage. The ISO/IEC 24029 series covers the assessment of the robustness of neural networks: Part 1 is an overview~\cite{iso2021artificial}, Part 2 describes formal methods~\cite{ISOIEC240292} and a planned Part 3 is expected to cover methodologies for the use of statistical methods~\cite{ISOIEC240293}. 
The  ISO/IEC 24029 Part 1 has been adopted as a European Standard. It provides a background of existing methods to assess the robustness of neural networks and explains the properties of statistical, formal and empirical methods. It states that "in practice, a combination of methods is used" and specifies that, "in the current state of the art, these methods are not used to directly assess robustness as a whole",  but instead "each targets complementary aspects of robustness, providing several partial indicators whose conjunction enables robustness assessment". The document remains primarily descriptive and does not specify how their combination is to be operationalised in practice. The report nevertheless recognises that robustness assessment depends on context. In the section on statistical methods is reported that "to evaluate the robustness of a model, a dataset that spans the distribution and the input conditions of interest for the target application is first established", but also acknowledges that "there is no general method to assess the relevance of a dataset". It further notes that, "based on the context, the task at hand and the nature of the data, some metrics are not always appropriate, as they can lead to irrelevant or misleading results" and while "there are also numerous task-specific metrics" it says that "their description is out of the scope of this document". It explicitly recognises that "the evaluation of robustness is case-specific, given that certain organisations or situations require different robustness goals and metrics to determine if a goal is met", but does not provide operational criteria for identifying the relevant context, selecting appropriate metrics, or determining how robustness goals should be specified across different applications. It therefore remains a high-level, horizontal account of available methods rather than a context-sensitive framework for their use.

In the CEN-CENELEC JTC21 document "AI Standards: Complete Detailed Overview"~\footnote{Accessible \href{https://jtc21.eu/wp-content/uploads/2025/06/CEN-CENELEC-JTC21-AI-Standards-Complete-Detailed-Overview.pdf}{\textbf{here}}} the programmed Part 3 of this series is expected to include adversarial robustness, distribution shift performance and statistical testing methods. The overview further states that the document intends to include “practical testing protocols and assessment criteria for different types of neural network applications", suggesting a more application-oriented approach than Part 1. However, from this description, Part 3 still appears to focus on testing methodologies rather than on a prior analysis of where relevant perturbations arise or how robustness requirements should vary across different domains and risk contexts. It may therefore be more practice-oriented, but not yet structured around a fully context-sensitive analysis. 
Also, robustness is still addressed exclusively for neural networks, treated as a uniform block without differentiation by architecture, application domain, or operational constraints, leaving other types of AI systems, such as supervised, unsupervised, generative and general-purpose models largely uncovered. 

Related standardisation documents point in the same direction. The "CEN-CENELEC Focus Group Report: Road Map on Artificial Intelligence"\footnote{Available on the \href{https://jtc21.eu/about/}{\textbf{JTC21 website}}}, aimed at mapping ongoing work and guiding future standardisation priorities, states that robustness is "case-specific", yet its proposed course of action treats robustness as an issue “already covered” in current standardisation activities. The Roadmap also acknowledges, in its discussion of simulation-based conformity assessment, unresolved questions concerning operating domains, including how they should be specified and whether they should be standardised at all. Taken together, these documents show that context sensitivity of robustness assessment is acknowledged, but not yet clearly operationalised. Although forthcoming standards may add further detail, the currently available documents do not indicate any clear structure for translating the recognised importance of context into standards that remain horizontal and broadly applicable across AI systems, even though, as discussed in Section~\ref{sec5.3}, systems developed in similar settings may still require very different robustness assessments.

To be effective, robustness standards should provide sufficiently clear requirements to enable providers to identify the relevant perturbations, select appropriate tests and metrics and determine acceptable thresholds in light of the intended context of application. This points to the need for a layered approach where horizontal standards are complemented with more detailed and context-sensitive specifications.

\section{Context-sensitive standardisation through a multi-layered approach} \label{sec7}

\begin{figure*}[t]
    \centering
    \includegraphics[width=\textwidth]{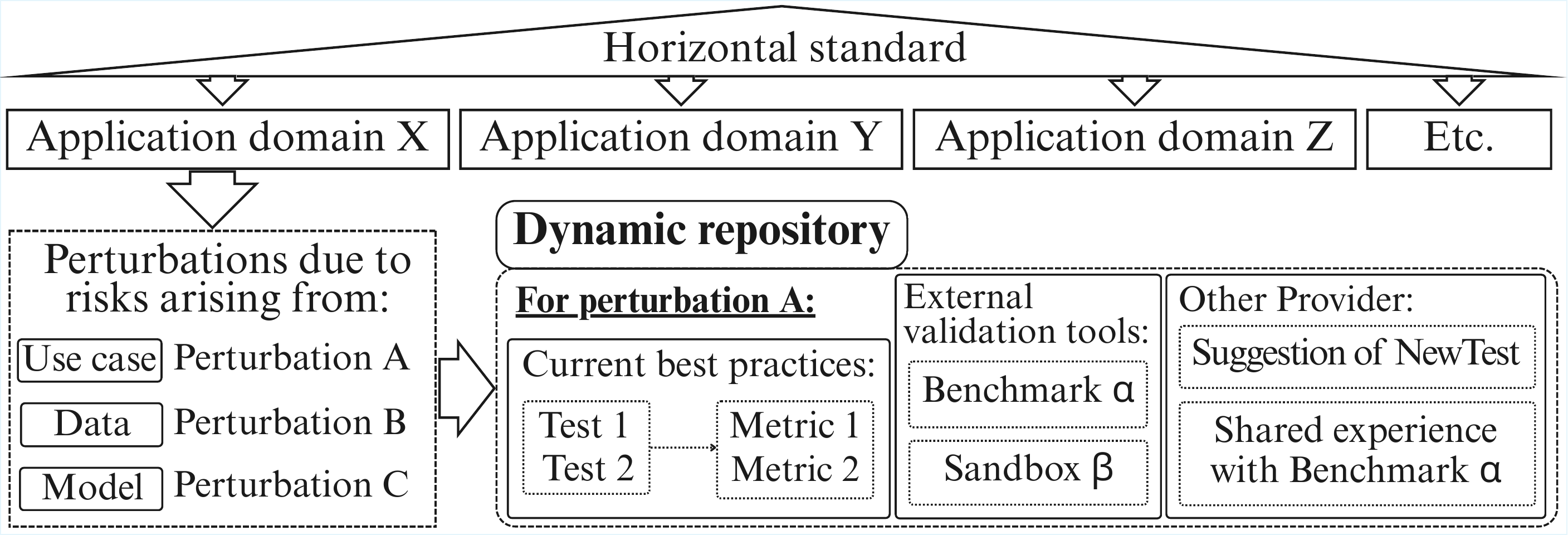}
    \caption{Multi-layered framework for context-sensitive robustness evaluation under the AI Act.}
    \label{fig:standard}

\end{figure*}

Horizontal standards provide a useful common core, but they must explicitly incorporate contextual hooks and minimal evidence requirements to adequately address the context-sensitive nature of robustness. 
Completely changing the current standardisation process may not be feasible, but its course can be adjusted. Horizontal standards could be designed as a starting point that establishes high-level principles, shared terminology and general best practices, while linking to more detailed standards that address domain-specific requirements. 
In this sense, context sensitivity could be incorporated through domain-specific standards, which would function as an incremental extension to the common horizontal baseline, thus building on planned deliverables rather than replacing them. A comparable logic can be found in the GDPR, which recognises sectoral codes of conduct as a way to specify how general rules apply in particular contexts and to support more effective implementation~\cite{gdpr2016}. 
A multi-layered approach would ensure coherence through common principles set in the horizontal documents and adaptability through tailored provisions for specific risks and contexts, improving auditability and allowing organisations to follow clear directions without having to interpret abstract requirements. 
In what follows, we propose a framework, summarised in Figure~\ref{fig:standard}, that directly addresses the shortcomings of current standards. It aims to reduce ambiguity by going beyond horizontal provisions and by identifying domain-specific risks across the AI lifecycle, guiding the choice of best practices for compliance with the AI Act context-sensitive requirements. 
Standardisation processes are typically long-term, iterative, shaped by negotiation among multiple stakeholders and take about three years from proposal to publication~\cite{iso_developing_standards}. Although this structure is essential to ensure stability and broad acceptability, it may also make standards less responsive to rapid technological change~\cite{wiarda2022responsible}. In fast-evolving fields such as AI, this can result in standards that do not fully reflect the most recent methods, risks, or practices, potentially leading to outdated compliance methodologies. 
For this reason, the framework proposes a context-sensitive and dynamic repository of recommended practices where stakeholders can propose new methods and experiences. This would enable the ESOs to track technological developments, update practices more efficiently and provide a shared platform that supports providers in both the design and validation of their systems.
This proposal also supports the objectives of the so-called HAS assessment process checklist~\footnote{A procedure to evaluate if a deliverable complies with standardisation requests and EU legal requirements. }. In particular, it better satisfies two central criteria: that the mandatory parts of a standard are written clearly, leaving no room for arbitrary choices, and that risk assessment is complete, with evidence that all relevant risks have been considered. The goal is not to replace existing processes, but to create a more structured framework where horizontal and context-sensitive standards complement each other, bringing clarity at the top level while respecting the diversity of needs on the ground.

\subsection{Domain-specific standardisation}

Robustness standards need to be domain-specific. AI is a broad umbrella term that includes a large number of subfields~\cite{ulnicane2021framing}. The AI Act identifies several high-risk application areas (see Section~\ref{sec5.1}) which provide a structured way to differentiate between domains for standardisation purposes. Adapting horizontal rules to such diverse applications is problematic, since each domain has unique characteristics~\cite{reddy2023navigating}. Different contexts require distinct ethical considerations and the requirements for AI systems change depending on the underlying technology and domain in which they are deployed. Domain-specific standardisation enables legal obligations to be translated into measurable criteria and better aligned with sectoral ethical priorities~\cite{gornet2024too}. For example, regulations that limit the use of patient data, suitable in general AI applications, may need to be modified in healthcare to allow the use of de-identified records~\cite{morley2020ethics} and ensure that the ethical principles that guide the industry, such as autonomy, beneficence, non-maleficence and justice, are respected~\cite{beauchamp1994principles}. Domain-specific standards do not entail fragmentation, as long as they are built on horizontal ones and would ensure completeness, alignment with general requirements and prevent ambiguities or arbitrary interpretations.

\subsection{Context-sensitive perturbation taxonomy}

\citet{JRC132833} highlight the need for guidance on how to set acceptable thresholds for different robustness methods, taking into account the context of use and risks of specific AI systems. Building on this, for each domain, there should be an exhaustive perturbation taxonomy. As discussed in Section~\ref{sec5}, perturbation identification is context-sensitive. Therefore, guidelines should map potential perturbations across the contextual drivers---use case, data, model---illustrate their root causes and suggest mitigation strategies. Ultimately, at each stage of implementation, providers determine the direction the system should take based on the perturbations identified up to that point. A similar proposal was advanced by \citet{schnitzer2023ai}, who developed a taxonomy of AI hazards linked to lifecycle stages. Although mainly focused on DNN-related hazards, it offers a solid foundation that could be adapted to robustness and to other requirements.

\subsection{Dynamic repository of best practices}

To address robustness tests, metrics and thresholds, it is necessary to refer to state-of-the-art practices, understood as widely accepted good practice at a given time~\cite{EU2023StandardisationAIAnnexes}. 
To support this, our framework proposes the development of a repository of best practices, maintained by the ESOs and informed by stakeholder input. 
The repository should link best practices and their limitations to the relevant risks to robustness—and to other requirements—that may arise during development and deployment, thereby supporting their evaluation and mitigation, including mitigation measures against biased outputs arising from feedback loops. Given the rapid technological change and the risk of standards obsolescence, the repository should allow providers to share experiences and propose new informative methodologies, complementing recommended practices until official updates are made. The proposed methodology would inform the ESOs of the areas that require updates, make the process faster, while shared experiences would help providers choose methods for their use cases and align system development with regulatory expectations. Since identifying all possible risks is unrealistic, standards will supply a range of technical options that providers can assess and implement in light of the purpose of their system, while the repository supports providers in selecting those most appropriate for their context. 
A similar initiative is the Artificial Intelligence Measurement and Evaluation (AIME)~\cite{team2021artificial}, launched by NIST, which develops metrics and methods for robustness and performance in areas such as language, vision and robotics. 
The program does not offer a centralised repository, but in collaboration with public and private sectors, NIST has developed a voluntary framework\footnote{ See \href{https://www.nist.gov/itl/ai-risk-management-framework}{\textbf{NIST AI Risk Management Framework}}} to manage AI-related risks for individuals, organisations and society.
The OECD has also developed a Catalogue of Tools \& Metrics for Trustworthy AI~\footnote{See \href{https://oecd.ai/en/catalogue/overview}{\textbf{OECD Catalogue}}}, which collects practices and metrics for fairness, transparency, explainability, robustness and safety. 
While it offers a valuable foundation, it is not up to date nor designed for context sensitivity, but it could serve as a reference for a dynamic European AI repository. 

\subsection{External validation}

Context sensitivity should also be extended to the external validation tools referenced in the AI Act: benchmarks and sandboxes. The repository should indicate which benchmarks are most suitable for different tasks and categories of high-risk systems. While a single benchmark can cover multiple systems (see Section~\ref{sec5.3}), interpretation of results must remain context-sensitive and should be interpreted with care~\cite{maier2018rankings}, therefore, sharing experiences would help providers better contextualise outcomes and highlight practical limitations. Moreover, benchmarks risk becoming obsolete within months~\cite{weidinger2025toward}, making the repository crucial for collecting and updating proposals for new benchmarks. Since benchmarks can fail to account for data or domain shifts, regulatory sandboxes introduced by the AI Act (Art. 57) provide a valuable complement to ensure that systems are fit for deployment. Sandboxes are controlled environments, established and supervised by national competent authorities, that facilitate technical testing under real or close-to-real conditions, providing resources, data and testing infrastructure~\cite{madiega2022artificial}. Yet, like best practices and benchmarks, sandboxes face context sensitivity due to the different needs of systems across sectors, where a one-size-fits-all solution is not feasible. As \citet{due2025sandboxing} note, stakeholders already stress the difficulty of designing a sandbox that can cover all participants’ needs, reinforcing the case for domain-specific sandboxes, especially for high-risk AI systems, even if the Act mandates horizontal national ones. Including context-sensitive sandboxes in the repository would guide use-case-specific testing and experience sharing. In this way, the repository becomes a practical tool to support both model design and model evaluation, enabling AI systems to be iteratively improved, compliant and aligned with context-specific requirements.

\section{Conclusions}

Robustness has a context-sensitive nature.
Its assessment should follow a structured approach that considers the system's intended use, the data available and the model architecture that would best fit the goal. The analysis of each of these contextual drivers narrows down the perturbations the system should be robust against and enables the selection of appropriate tests, evaluation and mitigation metrics. 
Shifts in contextual assumptions can lead to substantial differences in the assessment process. 
Since only harmonised standards confer presumption of conformity under the AI Act, this variability must be accounted for in their ongoing development. The current approach aims to design standards applicable to various types of AI systems across sectors. However, some requirements, such as robustness, are highly context-sensitive and horizontal standardisation may result in high-level directions that are operationally meaningless, leaving too much space for interpretation that could cause superficial compliance. 

Standards could follow a multi-layered approach that establishes high-level principles and points to domain-specific methodologies. To ensure robustness, a perturbation taxonomy is needed and it should be mapped into the system's lifecycle, guiding providers in the selection of the perturbations to test, evaluation methods and benchmarks. These should be listed in a repository designed for temporal adaptability to prevent technological obsolescence. 
The repository should also enable providers to share experiences and suggest new methods. Such contributions, while only informative, could complement recommended practices and inform future updates. In this way, robustness can be treated as a context-sensitive requirement, assessed through precise yet evolving standards that ensure compliance with the AI Act.

\bibliography{refnotes}

@techreport{JRC139430,
    author       = {Soler Garrido, Josep and De Nigris, Sarah and Bassani, Elias and Sanchez, Ignacio and Evas, Tatjana and André, Antoine-Alexandre and Boulangé, Thierry},
  title        = {Harmonised Standards for the European AI Act},
  year         = {2024},
  number       = {JRC139430},
  institution  = {Joint Research Centre (JRC), European Commission}
}

@techreport{JRC132833,
	
	author       = {Soler Garrido, Josep and Fano Yela, Delia and Panigutti, Cecilia and Junklewitz, Henrik and Hamon, Ronan and Evas, Tatjana and André, Antoine-Alexandre and Scalzo, Salvatore},
    title = {Analysis of the preliminary AI standardisation work plan in support of the AI Act},
    institution  = {Joint Research Centre (JRC), European Commission},
    year = {2023},
	isbn = {978-92-68-03924-3},
	doi = {10.2760/5847}
}

@misc{weidinger2025toward,
  title={Toward an evaluation science for generative AI systems},
  author={Weidinger, Laura and Raji, Deb and Wallach, Hanna and Mitchell, Margaret and Wang, Angelina and Salaudeen, Olawale and Bommasani, Rishi and Kapoor, Sayash and Ganguli, Deep and Koyejo, Sanmi and others},
  year={2025}
}

@misc{gornet2024too,
  title={Too broad to handle: can we" fix" harmonised standards on artificial intelligence by focusing on vertical sectors?},
  author={Gornet, M{\'e}lanie},
  year={2024},
  note = {HAL open archive}
}

@article{floridi2022capai,
  title={CapAI-A procedure for conducting conformity assessment of AI systems in line with the EU artificial intelligence act},
  author={Floridi, Luciano and Holweg, Matthias and Taddeo, Mariarosaria and Amaya, Javier and M{\"o}kander, Jakob and Wen, Yuni},
  journal={Available at SSRN 4064091},
  year={2022}
}

@techreport{iso2021artificial,
    author = {{CEN/CLC ISO/IEC/TR 24029-1}},
    title = {Artificial Intelligence {(AI)}—Assessment of the Robustness of Neural Networks — Part 1: Overview (ISO/IEC TR 24029-1:2021)},
    institution  = {European Committee for Standardisation and European Committee for Electrotechnical Standardisation},
    year = {2023}
}

@article{act2024artificial,
  title={Artificial Intelligence Act},
  author={{European Parliament and Council}},
  journal={EU’s Official Journal},
  volume={Regulation (EU) 2024/1689 of the European Parliament and of the Council of 13 June 2024 laying down harmonised rules on artificial intelligence (Artificial Intelligence Act)},
  year={2024}
}

@article{freiesleben2023beyond,
  title={Beyond generalization: a theory of robustness in machine learning},
  author={Freiesleben, Timo and Grote, Thomas},
  journal={Synthese},
  volume={202},
  number={109},
  year={2023},
  publisher={Springer}
}

@incollection{braiek2025machine,
  title={Machine learning robustness: A primer},
  author={Braiek, Houssem Ben and Khomh, Foutse},
  booktitle={Trustworthy AI in Medical Imaging},
  pages={37--71},
  year={2025},
  publisher={Elsevier}
}

@techreport{2024BEI2,
author = {{BS EN ISO/IEC 25059}},
title = {Software engineering — Systems and software 
Quality Requirements and Evaluation 
{(SQuaRE)} — Quality model for {AI} system},
institution  = {International Organisation for Standardisation (ISO)},
publisher = {British Standards Institute},
year = {2024}
}

@article{tocchetti2025ai,
  title={{AI} robustness: a human-centered perspective on technological challenges and opportunities},
  author={Tocchetti, Andrea and Corti, Lorenzo and Balayn, Agathe and Yurrita, Mireia and Lippmann, Philip and Brambilla, Marco and Yang, Jie},
  journal={ACM Computing Surveys},
  volume={57},
  number={6},
  pages={1--38},
  year={2025},
  publisher={ACM New York, NY}
}

@inproceedings{nobandegani2019robustness,
  title={On Robustness: An Undervalued Dimension of Human Rationality.},
  author={Nobandegani, Ardavan Salehi and da Silva Castanheira, Kevin and O'Donnell, Timothy and Shultz, Thomas R},
  booktitle={CogSci},
  pages={3327},
  year={2019}
}

@inproceedings{la2020assessing,
    title = {Assessing Robustness of Text Classification through Maximal Safe Radius Computation},
    author={La Malfa, Emanuele and Wu, Min and Laurenti, Luca and Wang, Benjie and Hartshorn, Anthony and Kwiatkowska, Marta},
    booktitle = {Findings of the Association for Computational Linguistics: EMNLP 2020},
    year = {2020},
    publisher = "Association for Computational Linguistics",
    pages = "2949--2968"
}

@misc{drenkow2021systematic,
  title={A systematic review of robustness in deep learning for computer vision: Mind the gap?},
  author={Drenkow, Nathan and Sani, Numair and Shpitser, Ilya and Unberath, Mathias},
  year={2021}
}

@inproceedings{singh2019boosting,
  title={Boosting robustness certification of neural networks},
  author={Singh, Gagandeep and Gehr, Timon and P{\"u}schel, Markus and Vechev, Martin},
  booktitle={International conference on learning representations},
  year={2019},
}

@article{yang2021enhancing,
  title={Enhancing robustness verification for deep neural networks via symbolic propagation},
  author={Yang, Pengfei and Li, Jianlin and Liu, Jiangchao and Huang, Cheng-Chao and Li, Renjue and Chen, Liqian and Huang, Xiaowei and Zhang, Lijun},
  journal={Formal Aspects of Computing},
  volume={33},
  number={3},
  pages={407--435},
  year={2021},
  publisher={Springer}
}

@inproceedings{dong2020benchmarking,
  title={Benchmarking adversarial robustness on image classification},
  author={Dong, Yinpeng and Fu, Qi-An and Yang, Xiao and Pang, Tianyu and Su, Hang and Xiao, Zihao and Zhu, Jun},
  booktitle={proceedings of the IEEE/CVF conference on computer vision and pattern recognition},
  pages={321--331},
  year={2020}
}

@book{chollet2021deep,
  title={Deep learning with Python},
  author={Chollet, Francois and Chollet, Fran{\c{c}}ois},
  year={2021},
  publisher={Simon and Schuster}
}

@book{geron2022hands,
  title={Hands-on machine learning with Scikit-Learn, Keras, and TensorFlow: Concepts, tools, and techniques to build intelligent systems},
  author={G{\'e}ron, Aur{\'e}lien},
  year={2022},
  publisher={" O'Reilly Media, Inc."}
}

@inproceedings{lassance2019structural,
  title={Structural robustness for deep learning architectures},
  author={Lassance, Carlos and Gripon, Vincent and Tang, Jian and Ortega, Antonio},
  booktitle={2019 IEEE Data Science Workshop (DSW)},
  pages={125--129},
  year={2019},
  organization={IEEE}
}

@inproceedings{croce2020robustbench,
  title={Robustbench: a standardized adversarial robustness benchmark},
  author={Croce, Francesco and Andriushchenko, Maksym and Sehwag, Vikash and Debenedetti, Edoardo and Flammarion, Nicolas and Chiang, Mung and Mittal, Prateek and Hein, Matthias},
  booktitle = {NeurIPS 2021 Datasets and Benchmarks Track (Round 2)},
  year={2020}
}

@book{quinonero2022dataset,
  title={Dataset shift in machine learning},
  author={Qui{\~n}onero-Candela, Joaquin and Sugiyama, Masashi and Schwaighofer, Anton and Lawrence, Neil D},
  year={2022},
  publisher={Mit Press}
}

@misc{brockman2016openai,
  title={Openai gym},
  author={Brockman, Greg and Cheung, Vicki and Pettersson, Ludwig and Schneider, Jonas and Schulman, John and Tang, Jie and Zaremba, Wojciech},
  doi = {10.48550/arXiv.1606.01540},
  year={2016}
}

@inproceedings{koh2021wilds,
  title={Wilds: A benchmark of in-the-wild distribution shifts},
  author={Koh, Pang Wei and Sagawa, Shiori and Marklund, Henrik and Xie, Sang Michael and Zhang, Marvin and Balsubramani, Akshay and Hu, Weihua and Yasunaga, Michihiro and Phillips, Richard Lanas and Gao, Irena and others},
  booktitle={International conference on machine learning},
  pages={5637--5664},
  year={2021},
  organization={PMLR}
}

@article{ghaffari2022adversarial,
   author       = {Ghaffari Laleh, Narmin and Truhn, Daniel and Veldhuizen, Gregory Patrick and Han, Tianyu and van Treeck, Marko and Buelow, Roman D. and Langer, Rupert and Dislich, Bastian and Boor, Peter and Schulz, Volkmar and Kather, Jakob Nikolas},
  title        = {Adversarial Attacks and Adversarial Robustness in Computational Pathology},
  journal      = {Nature Communications},
  volume       = {13},
  number       = {1},
  pages        = {5711},
  year         = {2022},
  publisher    = {Nature Publishing Group}
}

@article{sathitratanacheewin2020deep,
  title={Deep learning for automated classification of tuberculosis-related chest X-Ray: dataset distribution shift limits diagnostic performance generalizability},
  author={Sathitratanacheewin, Seelwan and Sunanta, Panasun and Pongpirul, Krit},
  journal={Heliyon},
  volume={6},
  number={8},
  year={2020},
  publisher={Elsevier}
}

@inproceedings{szegedy2013intriguing,
  title={Intriguing properties of neural networks},
  author={Szegedy, Christian and Zaremba, Wojciech and Sutskever, Ilya and Bruna, Joan and Erhan, Dumitru and Goodfellow, Ian and Fergus, Rob},
  booktitle    = {2nd International Conference on Learning Representations, {ICLR} 2014,
                  Banff, AB, Canada, April 14-16, 2014, Conference Track Proceedings},
  year={2013}
}

@article{goodfellow2014generative,
  title={Generative adversarial nets},
  author={Goodfellow, Ian J and Pouget-Abadie, Jean and Mirza, Mehdi and Xu, Bing and Warde-Farley, David and Ozair, Sherjil and Courville, Aaron and Bengio, Yoshua},
  journal={Advances in neural information processing systems},
  volume={27},
  year={2014}
}

@inproceedings{tian2018deeptest,
  title={Deeptest: Automated testing of deep-neural-network-driven autonomous cars},
  author={Tian, Yuchi and Pei, Kexin and Jana, Suman and Ray, Baishakhi},
  booktitle={Proceedings of the 40th international conference on software engineering},
  pages={303--314},
  year={2018}
}

@article{zhou2019metamorphic,
  title={Metamorphic testing of driverless cars},
  author={Zhou, Zhi Quan and Sun, Liqun},
  journal={Communications of the ACM},
  volume={62},
  number={3},
  pages={61--67},
  year={2019},
  publisher={ACM New York, NY, USA}
}

@inproceedings{shen2018munn,
  author       = {Shen, Weijun and Wan, Jun and Chen, Zhenyu},
  title        = {{MuNN}: Mutation Analysis of Neural Networks},
  booktitle    = {2018 IEEE International Conference on Software Quality, Reliability and Security Companion (QRS-C)},
  pages        = {108--115},
  year         = {2018},
  publisher    = {IEEE}
}

@inproceedings{ma2018deepmutation,
  title={Deepmutation: Mutation testing of deep learning systems},
  author={Ma, Lei and Zhang, Fuyuan and Sun, Jiyuan and Xue, Minhui and Li, Bo and Juefei-Xu, Felix and Xie, Chao and Li, Li and Liu, Yang and Zhao, Jianjun and others},
  booktitle={2018 IEEE 29th international symposium on software reliability engineering (ISSRE)},
  pages={100--111},
  year={2018},
  organization={IEEE}
}

@misc{ness2024medfuzz,
  title={Medfuzz: Exploring the robustness of large language models in medical question answering},
  author={Ness, Robert Osazuwa and Matton, Katie and Helm, Hayden and Zhang, Sheng and Bajwa, Junaid and Priebe, Carey E and Horvitz, Eric},
  year={2024},
 note = {Submitted to ICLR 2025}
}

@misc{christopher2008introduction,
  title={Introduction to information retrieval},
   author       = {Manning, Christopher D. and Raghavan, Prabhakar and Schütze, Hinrich},
  year={2008},
  publisher={Cambridge University Press}
}

@inproceedings{tsipras2018robustness,
  title={Robustness may be at odds with accuracy},
  author={Tsipras, Dimitris and Santurkar, Shibani and Engstrom, Logan and Turner, Alexander and Madry, Aleksander},
  booktitle    = {7th International Conference on Learning Representations, {ICLR} 2019,
                  New Orleans, LA, USA, May 6-9, 2019},
  publisher    = {OpenReview.net},
  year         = {2019}
}

@misc{deng2019architecture,
  title={Architecture selection via the trade-off between accuracy and robustness},
  author={Deng, Zhun and Dwork, Cynthia and Wang, Jialiang and Zhao, Yao},
  year={2019}
}

@inproceedings{juraev2024impact,
  title={Impact of architectural modifications on deep learning adversarial robustness},
  author={Juraev, Firuz and Abuhamad, Mohammed and Woo, Simon S and Thiruvathukal, George K and Abuhmed, Tamer},
booktitle={2024 Silicon Valley Cybersecurity Conference (SVCC)},
pages={1--7},
doi={10.1109/SVCC61185.2024.10637362},
  year={2024}
}

@inproceedings{
zheng2021graph,
title={Graph Robustness Benchmark: Benchmarking the Adversarial Robustness of Graph Machine Learning},
author={Qinkai Zheng and Xu Zou and Yuxiao Dong and Yukuo Cen and Da Yin and Jiarong Xu and Yang Yang and Jie Tang},
booktitle={Thirty-fifth Conference on Neural Information Processing Systems Datasets and Benchmarks Track (Round 2)},
year={2021}
}

@article{balendran2025scoping,
  title={A scoping review of robustness concepts for machine learning in healthcare},
  author={Balendran, Alan and Beji, C{\'e}line and Bouvier, Florie and Khalifa, Ottavio and Evgeniou, Theodoros and Ravaud, Philippe and Porcher, Rapha{\"e}l},
  journal={npj Digital Medicine},
  volume={8},
  number={1},
  pages={38},
  year={2025},
  publisher={Nature Publishing Group UK London}
}

@article{madiega2022artificial,
  title={Artificial intelligence act and regulatory sandboxes},
  author={Madiega, Tambiama and Van De Pol, Anne Louise},
  journal={European Parliamentary Research Service},
  volume={6},
  year={2022}
}

@article{ulnicane2021framing,
  title={Framing governance for a contested emerging technology: insights from AI policy},
  author={Ulnicane, Inga and Knight, William and Leach, Tonii and Stahl, Bernd Carsten and Wanjiku, Winter-Gladys},
  journal={Policy and Society},
  volume={40},
  number={2},
  pages={158--177},
  year={2021},
  publisher={Oxford University Press}
}

@article{reddy2023navigating,
  title={Navigating the AI revolution: the case for precise regulation in health care},
  author={Reddy, Sandeep},
  journal={Journal of medical Internet research},
  volume={25},
  pages={e49989},
  year={2023},
  publisher={JMIR Publications Toronto, Canada}
}

@book{beauchamp1994principles,
  title={Principles of biomedical ethics},
  author={Beauchamp, Tom L and Childress, James F},
  year={1994},
  publisher={Edicoes Loyola}
}

@article{morley2020ethics,
  title={The ethics of AI in health care: a mapping review},
  author = {Jessica Morley and Caio C.V. Machado and Christopher Burr and Josh Cowls and Indra Joshi and Mariarosaria Taddeo and Luciano Floridi},
  journal={Social science \& medicine},
  volume={260},
  pages={113172},
  year={2020},
  publisher={Elsevier}
}

@inproceedings{schnitzer2023ai,
  title={AI Hazard Management: A framework for the systematic management of root causes for AI risks},
  author={Schnitzer, Ronald and Hapfelmeier, Andreas and Gaube, Sven and Zillner, Sonja},
  booktitle={International Conference on Frontiers of Artificial Intelligence, Ethics, and Multidisciplinary Applications},
  pages={359--375},
  year={2023},
  organization={Springer}
}

@article{team2021artificial,
  title={Artificial intelligence measurement and evaluation at the national institute of standards and technology},
  author={Team, AIME Planning},
  journal={National Institute of Standards and Technology},
  year={2021}
}

@article{maier2018rankings,
  title={Why rankings of biomedical image analysis competitions should be interpreted with care},
  author={Maier-Hein, Lena and Eisenmann, Matthias and Reinke, Annika and Onogur, Sinan and Stankovic, Marko and Scholz, Patrick and Arbel, Tal and Bogunovic, Hrvoje and Bradley, Andrew P and Carass, Aaron and others},
  journal={Nature communications},
  volume={9},
  number={1},
  pages={5217},
  year={2018},
  publisher={Nature Publishing Group UK London}
}

@article{doerr2018iohprofiler,
  title={IOHprofiler: A benchmarking and profiling tool for iterative optimization heuristics},
  author={Doerr, Carola and Wang, Hao and Ye, Furong and Van Rijn, Sander and B{\"a}ck, Thomas},
  journal      = {CoRR},
  volume       = {abs/1810.05281},
  year         = {2018}
}

@article{wang2022iohanalyzer,
  title={IOHanalyzer: Detailed performance analyses for iterative optimization heuristics},
  author={Wang, Hao and Vermetten, Diederick and Ye, Furong and Doerr, Carola and B{\"a}ck, Thomas},
  journal={ACM Transactions on Evolutionary Learning and Optimization},
  volume={2},
  number={1},
  pages={1--29},
  year={2022},
  publisher={ACM New York, NY}
}

@misc{EU2023StandardisationAI,
 author       = {{Standardisation Request M/593}}, 
title        = {{Commission Implementing Decision of 22 May 2023 on a standardisation request to the European Committee for Standardisation and the European Committee for Electrotechnical Standardisation in support of Union policy on artificial intelligence}},
  howpublished = {Official Journal of the European Union, C/2023/3259},
  year         = {2023},
  note         = {Available at: \url{https://ec.europa.eu/transparency/documents-register/detail?ref=C(2023)3215&lang=en}},
}

@misc{EU2023StandardisationAIAnnexes,
  author       = {{Standardisation request M/593 Annexes}},
  title        = {{Annexes to the Commission Implementing Decision of 22 May 2023 on a standardisation request to the European Committee for Standardisation and the European Committee for Electrotechnical Standardisation in support of Union policy on artificial intelligence}},
  howpublished = {Official Journal of the European Union, C/2023/3259},
  year         = {2023}
}

@misc{due2025sandboxing,
  author       =   {Due, Stig A. and Shah, Hira and Moraes, Thiago and Genicot, Nathan and Canter, Martin},
  title        = {Sandboxing Artificial Intelligence: Balancing Innovation, Regulation, and Stakeholder Needs},
  year         = {2025},
  howpublished = {\url{https://www.fari.brussels/research-and-innovation/publication/sandboxing-artificial-intelligence}},
  note         = {FARI Institute}
}

@article{javed2025robustness,
  title={Robustness in deep learning models for medical diagnostics: security and adversarial challenges towards robust {AI} applications},
   author       = {Javed, Haseeb and El-Sappagh, Shaker and Abuhmed, Tamer},
  journal={Artificial Intelligence Review},
  volume={58},
  number={12},
  year={2025},
  doi={10.1007/s10462-024-11005-9},
  publisher={Springer}
}

@misc{dong2021data,
  title={Data quality matters for adversarial training: An empirical study},
  author={Dong, Chengyu and Liu, Liyuan and Shang, Jingbo},
  year={2021}
}

@misc{qiu2022benchmarking,
  title={Benchmarking robustness of multimodal image-text models under distribution shift},
  author={Qiu, Jielin and Zhu, Yi and Shi, Xingjian and Wenzel, Florian and Tang, Zhiqiang and Zhao, Ding and Li, Bo and Li, Mu},
  year={2022}
}

@article{roy2023hyperparameter,
  title={Hyperparameter optimization for deep neural network models: a comprehensive study on methods and techniques},
  author={Roy, Sunita and Mehera, Ranjan and Pal, Rajat Kumar and Bandyopadhyay, Samir Kumar},
  journal={Innovations in Systems and Software Engineering},
  pages={1--12},
  year={2023},
  publisher={Springer}
}

@article{arnold2024role,
  title={The role of hyperparameters in machine learning models and how to tune them},
  author={Arnold, Christian and Biedebach, Luka and K{\"u}pfer, Andreas and Neunhoeffer, Marcel},
  journal={Political Science Research and Methods},
  volume={12},
  number={4},
  pages={841--848},
  year={2024},
  publisher={Cambridge University Press}
}

@misc{zhang2025evaluating,
  title={Evaluating and Improving Robustness in Large Language Models: A Survey and Future Directions},
  author={Zhang, Kun and Wu, Le and Yu, Kui and Lv, Guangyi and Zhang, Dacao},
  year={2025}
}

@misc{li2023robustness,
  title={Robustness May be More Brittle than We Think under Different Degrees of Distribution Shifts},
  author={Li, Kaican and Zhang, Yifan and Hong, Lanqing and Li, Zhenguo and Zhang, Nevin L},
  year={2023}
}

@inproceedings{hendrycks2021many,
  title={The many faces of robustness: A critical analysis of out-of-distribution generalization},
  author={Hendrycks, Dan and Basart, Steven and Mu, Norman and Kadavath, Saurav and Wang, Frank and Dorundo, Evan and Desai, Rahul and Zhu, Tyler and Parajuli, Samyak and Guo, Mike and others},
  booktitle={Proceedings of the IEEE/CVF international conference on computer vision},
  pages={8340--8349},
  year={2021}
}

@inproceedings{farquhar2022out,
  title={What'out-of-distribution'is and is not},
  author={Farquhar, Sebastian and Gal, Yarin},
  booktitle={Neurips ml safety workshop},
  year={2022}
}

@article{laux2024trustworthy,
  title={Trustworthy artificial intelligence and the European Union AI act: On the conflation of trustworthiness and acceptability of risk},
  author={Laux, Johann and Wachter, Sandra and Mittelstadt, Brent},
  journal={Regulation \& Governance},
  volume={18},
  number={1},
  pages={3--32},
  year={2024},
  publisher={Wiley Online Library}
}

@inproceedings{nolte2025robustness,
  title={Robustness and cybersecurity in the EU Artificial Intelligence Act},
  author={Nolte, Henrik and Rateike, Miriam and Finck, Mich{\`e}le},
  booktitle={Proceedings of the 2025 ACM Conference on Fairness, Accountability, and Transparency},
  pages={283--295},
  year={2025}
}

@misc{deck2024implications,
  title={Implications of the AI act for non-discrimination law and algorithmic fairness},
  author={Deck, Luca and M{\"u}ller, Jan-Laurin and Braun, Conradin and Zipperling, Domenique and K{\"u}hl, Niklas},
  year={2024}
}

@book{finck2026eu,
  title={The EU Artificial Intelligence Act: A Commentary},
  author={Finck, Mich{\`e}le},
  year={2026},
  publisher={Oxford University Press}
}

@book{european2003joint,
  title={Joint Practical Guide of the European Parliament, the Council and the Commission for Persons Involved in the Drafting of Legislation Within the Community Institutions},
  author={European Parliament and Council of the European Union},
  year={2003},
  publisher={Luxembourg: Office for Official Publications of the European Communities}
}

@misc{EUCommission2026UnderstandingStandardisationAIAct,
  author       = {{European Commission}},
  title        = {Understanding the Standardisation of the AI Act},
  year         = {2026},
  howpublished = {\url{https://digital-strategy.ec.europa.eu/en/faqs/understanding-standardisation-ai-act}},
  note         = {Shaping Europe’s Digital Future, Directorate-General for Communications Networks, Content and Technology, last updated 10 March 2026, accessed 22 March 2026}
}

@misc{ISOIEC240292,
  author       = {{International Organization for Standardization}},
  title        = {{ISO/IEC 24029-2:2023} --- Artificial Intelligence {(AI)} --- Assessment of the robustness of neural networks --- Part 2: Methodology for the use of formal methods},
  year         = {2023}
}

@misc{ISOIEC240293,
  author       = {{International Organization for Standardization}},
  title        = {{ISO/IEC DIS 24029-3} --- Artificial Intelligence {(AI)} --- Assessment of the robustness of neural networks --- Part 3: Methodology for the use of statistical methods},
  year         = {2026},
  note         = {Draft under development.}
}

@misc{iso_developing_standards,
  author       = {{International Organization for Standardization}},
  title        = {Developing standards},
  url          = {https://www.iso.org/developing-standards.html},
  note         = {ISO},
}

@article{wiarda2022responsible,
  title={Responsible innovation and de jure standardisation: An in-depth exploration of moral motives, barriers, and facilitators},
  author={Wiarda, Martijn and van de Kaa, Geerten and Doorn, Neelke and Yaghmaei, Emad},
  journal={Science and Engineering Ethics},
  volume={28},
  number={6},
  pages={65},
  year={2022},
  publisher={Springer}
}

@book{malgieri2026eu,
  title={The EU Artificial Intelligence Act: A Thematic Commentary},
  author={Malgieri, G. and Fuster, G.G. and Mantelero, A. and Zanfir-Fortuna, G.},
  isbn={9781509988600},
  year={2026},
  publisher={Bloomsbury Academic}
}

@misc{gdpr2016,
  author       = {{European Parliament and Council of the European Union}},
  title        = {Regulation (EU) 2016/679 of the European Parliament and of the Council of 27 April 2016 on the protection of natural persons with regard to the processing of personal data and on the free movement of such data, and repealing Directive 95/46/EC (General Data Protection Regulation)},
  year         = {2016},
  howpublished = {Official Journal of the European Union, L 119, 4 May 2016, pp. 1--88}
}

@misc{wilmet2021comparison,
  title={A comparison of supervised and unsupervised deep learning methods for anomaly detection in images},
  author={Wilmet, Vincent and Verma, Sauraj and Redl, Tabea and Sandaker, H{\aa}kon and Li, Zhenning},
  year={2021}
}

\end{document}